\documentclass[acmtog]{acmart}
\AtBeginDocument{%
  \providecommand\BibTeX{{%
    \normalfont B\kern-0.5em{\scshape i\kern-0.25em b}\kern-0.8em\TeX}}}

\setcopyright{acmcopyright}
\copyrightyear{2023}
\acmYear{2023}

\definecolor{lblue}{HTML}{A6CEE3}
\definecolor{lgreen}{HTML}{B2DF8A}
\definecolor{lred}{HTML}{FB9A99}
\definecolor{lorange}{HTML}{FDBF6F}
\definecolor{mblue}{HTML}{80B1D3}
\definecolor{mgreen}{HTML}{B3DE69}
\definecolor{mred}{HTML}{FB8072}
\definecolor{morange}{HTML}{FDB462}
\definecolor{blue}{HTML}{1F78B4}
\definecolor{green}{HTML}{33A02C}
\definecolor{red}{HTML}{E31A1C}
\definecolor{orange}{HTML}{FF7F00}
\definecolor{dblue}{HTML}{08519C}
\definecolor{dgreen}{HTML}{006D2C}
\definecolor{dorange}{HTML}{EC7014}

\newcommand{\header}[1]{\vspace{1mm}\noindent\textbf{#1.}}

\AtBeginDocument{%
  \providecommand\BibTeX{{%
    \normalfont B\kern-0.5em{\scshape i\kern-0.25em b}\kern-0.8em\TeX}}}

\usepackage{acronym}
\usepackage{natbib}
\usepackage{numprint}
\usepackage[inline]{enumitem}
\usepackage{natbib}
\usepackage{bbm}
\usepackage{breakcites}

\usepackage{multicol}
\usepackage{multirow}

\acrodef{SOTA}{state-of-the-art}
\acrodef{IR}{information retrieval}
\acrodef{CNN}{convolutional neural network}
\acrodef{CNNs}{convolutional neural networks}
\acrodef{MFCCs}{mel frequency capstral coefficients}
\acrodef{LSTM}{long short-term memory}
\acrodef{MF}{matrix factorisation}
\acrodef{CLIP}{Contrastive Language-Image Pretraining}
\acrodef{MSD}{Million Song Dataset}
\acrodef{AUC}{area under the curve}
\acrodef{CBR}{content-based recommender}
\acrodef{CF}{collaborative filtering}
\acrodef{CL}{contrastive learning}
\acrodef{LMS}{log mel spectogram}

\newcommand{\ourmodel}{$CL_{BASE}$}

\newcommand\sL{\ensuremath{\mathcal{L}}}

\newcommand\bq{\ensuremath{\mathbf{q}}}

\newcommand\bx{\ensuremath{\mathbf{x}}}

\usepackage{color}

\begin{document}

\title{Towards Contrastive Learning in Music Video Domain}

\author{Karel Veldkamp, Mariya Hendriksen, Zoltán Szlávik, Alexander Keijser}

\begin{abstract}
  Contrastive learning is a powerful way of learning multimodal representations across various domains such as image-caption retrieval and audio-visual representation learning. In this work, we investigate if these findings generalize to the domain of music videos.
    Specifically, we create a dual encoder for the audio and video modalities and train it using a bidirectional contrastive loss. For the experiments, we use an industry dataset containing \numprint{550000} music videos as well as the public \acl{MSD}, and evaluate the quality of learned representations on the downstream tasks of music tagging and genre classification. Our results indicate that pre-trained networks without contrastive fine-tuning outperform our contrastive learning approach when evaluated on both tasks.
    To gain a better understanding of the reasons contrastive learning was not successful for music videos, we perform a qualitative analysis of the learned representations, revealing why contrastive learning might have difficulties uniting embeddings from two modalities. Based on these findings, we outline possible directions for future work. To facilitate the reproducibility of our results, we share our code and the pre-trained model.
\end{abstract}

\begin{CCSXML}
<ccs2012>
 <concept>
  <concept_id>10010520.10010553.10010562</concept_id>
  <concept_desc>Computer systems organization~Embedded systems</concept_desc>
  <concept_significance>500</concept_significance>
 </concept>
 <concept>
  <concept_id>10010520.10010575.10010755</concept_id>
  <concept_desc>Computer systems organization~Redundancy</concept_desc>
  <concept_significance>300</concept_significance>
 </concept>
 <concept>
  <concept_id>10010520.10010553.10010554</concept_id>
  <concept_desc>Computer systems organization~Robotics</concept_desc>
  <concept_significance>100</concept_significance>
 </concept>
 <concept>
  <concept_id>10003033.10003083.10003095</concept_id>
  <concept_desc>Networks~Network reliability</concept_desc>
  <concept_significance>100</concept_significance>
 </concept>
</ccs2012>
\end{CCSXML}

\keywords{Music video recommendation, contrastive learning}

\maketitle

% !TEX root = ../main.tex

\section{Introduction}
One of the most important challenges in music video recommendation is that music videos are most relevant shortly after their release, resulting in the need for an approach that works well for new items. A class of methods that are known to work well for new items are \ac{CBR} systems. However, these methods rely on the quality of item representations \citep{volkovs2017content, zhu2019addressing}. In the use case of music videos, these representations should contain information from both the audio and video domains.  In contrast to music representation learning \cite{kim2020one}, the topic of learning music video representations is relatively unexplored.

% A popular alternative to \ac{CB} methods is \ac{CF} \cite{volkovs2017content, zhu2019addressing}. However, \ac{CF} methods are based on historical usage data, meaning that they have issues dealing with new items. This is known as the cold-start problem. Since music videos are released often and are usually most relevant shortly after their release, \ac{CB} recommendation seems most appropriate for the task. Additionally, another downside of popular \ac{CF} methods is that they are known to exhibit a bias for more popular items, which can hurt user experience for users with a more alternative taste \cite{boratto2021connecting}.

One method for multimodal representation learning that has been proven to be powerful across various domains is \ac{CL}\cite{chai2022deep, shi2023self}. This class of methods uses pairs of data points from different modalities to learn multimodal representations, thereby avoiding the need for labels. The most renowned application of this methodology was in the text-image domain\citep{radford2021learning, hendriksen2022multimodal}, but it has also been applied in the audio-video domain \cite{wang2021multimodal, ma2020active}. In this paper, we examine whether \ac{CL} can successfully be generalized to music videos.

This research was carried out in collaboration with a music video streaming platform. The platform provided a private dataset of \numprint{555000} music videos. Although this company dataset provides us with the unique opportunity to test \ac{SOTA} methods on a large collection of real music videos, it also means that we can not share our training data publicly, since it largely consists of copyrighted music videos. However, to keep this research as reproducible as possible, we evaluate our trained model on a subsection of the publicly available \ac{MSD} that overlaps with the company dataset. Additionally, our code and trained models are made publicly available, along with a description of the company dataset and a comparison to the public \ac{MSD}.

\header{Research questions}
We devise a method that learns audio-visual representations for music videos based on their content using \ac{CL}. We aim to answer the following research questions:
\begin{enumerate*}[label=(RQ\arabic*)]
    \item Can \ac{CL} of audio-video representations be used to improve the quality of music video representations relative to existing models when evaluated on the downstream tasks of genre classification and music tagging?
    \item Can our audio-visual representation be used to calculate music-video similarity that is in line with the subjective human judgment of similarity? 
\end{enumerate*}

\header{Main findings}
We discover that our training procedure was not able to pull embeddings for corresponding pairs of audio and video together. Consequently, the contrastive model is outperformed by the baselines on the downstream tasks, indicating that our approach was not able to generalize contrastive learning to the music video setting. Additionally, we find that human judgment is not in line with the similarity of representations.

To better understand these outcomes, we perform a qualitative analysis of the learned representations. Specifically, we perform an error analysis by looking at similarities and dissimilarities of music-video clips that were retrieved based on various seed items. This reveals a potential explanation for the failure of the contrastive loss function to pull embeddings from the two modalities together. 

\noindent
\header{Contributions}
The principal contributions of our research are the following:
\begin{enumerate*}[label=(\arabic*)]
	\item We adopt a \ac{CL} approach to learn music video representations for recommendations and evaluate its ability to generalize to the music video domain.
	\item We provide a quantitative evaluation of the learned representations on two downstream tasks of genre classification and music tagging.
	\item A qualitative evaluation of the representations based on item similarity.
	\item To facilitate the reproducibility of our work, we provide the code of our experiments, a pre-trained model, and a description of the company dataset on 
	\href{https://github.com/KarelVeldkamp/Multimodal-Musicvideo-Representation}{GitHub}\footnote{\url{https://github.com/KarelVeldkamp/Multimodal-Musicvideo-Representation}}.
\end{enumerate*}

 % !TEX root = ../main.tex
 
\section{Related work}
% \header{Unimodal approaches} A popular approach to modeling in this field is to train a \ac{CNN} that convolves along the time and frequency domain of the \ac{LMS}. This approach was used for various music-related tasks\cite{choi2016automatic, akbari2021vatt, won2020evaluation}. In a comparative study of models trained on the task of music tagging, \citet{won2020evaluation} found that a specific \ac{CNN} called \emph{musicnn}\cite{pons2019musicnn} could get good results with fewer data compared to different \ac{CNN}s and other \ac{SOTA} models.
% % The \emph{musicnn} model takes as input a log mel spectrogram with 96 bands over three seconds represented as 187-time points. 

% \ac{CNN}s are also a popular approach in the video domain. \citet{tran2018closer} attain \ac{SOTA} performance on the task of action recognition with their \emph{R3D} model, which convolves along the temporal dimension as well as the two spatial dimensions. Additionally, the authors show that the three-dimensional convolutions can be separated into separate 2D convolutions for space, and 1D convolutions for time. This \emph{R(2+1)D} model reduces the complexity of the model significantly, without degrading performance.

\header{Multimodal learning}
The main challenge for multimodal learning is that feature vectors for different modalities encode different types of information, which is referred to as the heterogeneity gap\cite{guo2019deep, hendriksen2023scene}. A traditional approach to deal with this problem is to project the different modalities into a single subspace while retaining information from the different modalities\cite{baltruvsaitis2018multimodal}. This is often done in a supervised \cite{jiang2017exploiting, nojavanasghari2016deep, ortega2019multimodal} or unsupervised \cite{ngiam2011multimodal, silberer2014learning, goei2021tackling} manner. One recent approach gaining traction is \ac{CL}\cite{chai2022deep}. 

\header{Contrastive learning}
\label{contrastivelearning}
\ac{CL} methods learn from pairs of samples by pulling representations closer together for positive pairs of samples\citep{le2020contrastive}. In a unimodal setting pairs are created using augmentation\cite{chen2020simple, spijkervet2021contrastive, lorre2020temporal, shi2023rethink}, whereas in a multimodal setting, positive pairs are matching samples from different modalities. A renowned study in this field introduced \ac{CLIP}\cite{radford2021learning}. \ac{CLIP} learned a multimodal embedding space for images and text using \ac{CL} on a large dataset of image-caption pairs. The model attains performance on a variety of vision-language tasks\cite{hendriksen2022unimodal}. The approach works by using separate modality-specific encoder networks, which are initialized using a pre-trained network for the relevant modality. A projection head is added on top, which consists of extra feedforward layers with the same output dimensions across modalities. The network is trained contrastively, by maximizing the similarity of the output layers for image-text pairs that belong together and minimizing similarity for pairs that do not. This seemingly simple training task results in informative representations of the input data, evident from the high zero-shot performance on various tasks. Several studies have been published using variations of this method including both the audio and video domains\citep{guzhov2021audioclip, wang2021multimodal, ma2020active, akbari2021vatt, hendriksen2022extending}.

Although earlier research has shown the potential of \ac{CL} in the audio-video domain, it has not previously been applied for music video domain. Prior work mainly focused on the kinetics human action dataset\cite{kay2017kinetics}, the AudioSet dataset\cite{gemmeke2017audio}, and the HowTo100M dataset\cite{miech2019howto100m}. The relationship between a song and its video is more abstract and more challenging to learn.

\section{Approach}

\begin{figure}[h]
\centering
\includegraphics[width=8.2cm]{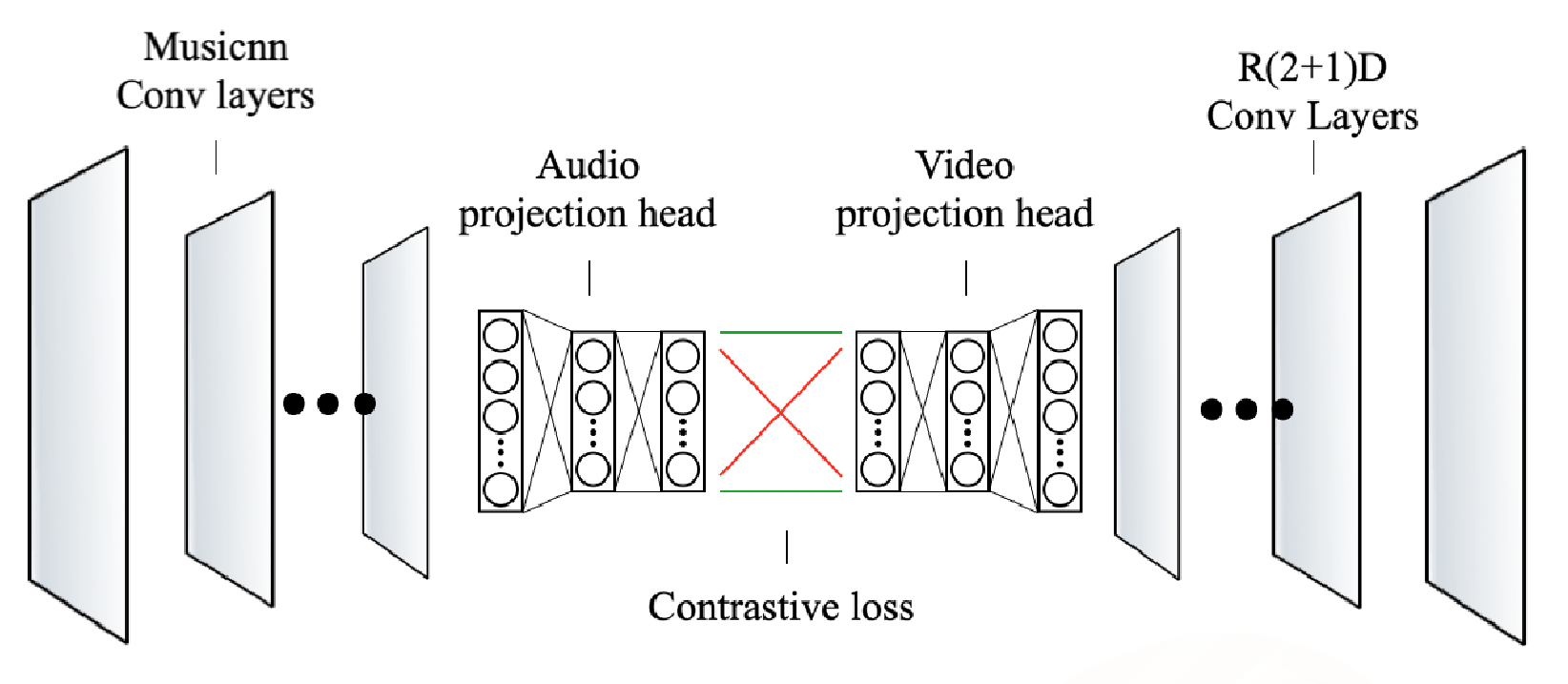}
\caption{Model architecture, comprised of pre-trained convolutional layers of the
\emph{musicnn} and R(2+1)D networks, and a dense projection head for each modality. We freeze the convolutional layers and train the projection heads.}
\label{architecture}
\end{figure}

Figure\ref{architecture} presents an overview of our proposed architecture.

\header{Audio encoding pipeline} Audio is encoded using the encoder network $f_a$ and a projection head $g_a$. We use the pre-trained \emph{musicnn} network as encoder\cite{pons2019musicnn}, freezing the weights during training.  The projection head consists of two dense feed-forward layers. We use a hidden layer size of 512 and an embedding size of 256. These embedding sizes are relatively small compared to some other multimodal networks\cite{radford2021learning, faghri2017vse++} because we have relatively little training data. The hidden layer uses a ReLU activation function and is trained using a dropout rate of 0.3. The embedding layer uses a sigmoid activation function. 
    
\header{Video encoding pipeline} as video encoder $f_v$, we use the (2+1)D \ac{CNN} from\citet{tran2018closer}. This encoder is initialized with the pre-trained weights and is frozen during training. We use the same projection head architecture for the video encoder as we did for the audio encoder. 
We also experiment with several variations to the architecture of our two projection heads, which are discussed in Section\ref{experimentalsetup}.

\header{Loss Function}
\label{loss}
To train the dense layers of the two projection heads, we use the contrastive loss function from simCLR\cite{chen2020simple}, adapted to the audio-video scenario. For a batch of $N$ music videos, video-to-audio and audio-to-video losses for a positive pair of audio embedding $a_j$ and video embedding $v_j$ are defined respectively as:

\begin{equation}\label{lossfunction}
l^{(v\rightarrow a)}_j = -log \frac{\exp(f_{sim}(a_j,v_j)/\tau)}{\sum_{k=1}^{N} \mathbbm{1}_{[k\neq j]} \exp(f_{sim}(a_j,v_k)/\tau)}.
\end{equation}
\begin{equation}\label{lossfunction}
l^{(a\rightarrow v)}_j = -log \frac{\exp(f_{sim}(v_j,a_j)/\tau)}{\sum_{k=1}^{N} \mathbbm{1}_{[k\neq j]} \exp(f_{sim}(v_j,a_k)/\tau)}.
\end{equation}

Here $f_{sim}(x,y)$ is the cosine similarity function: $ f_{sim}(\bq, \bx) = \frac{\bq}{||\bq||} \frac{\bx}{||\bx||}$; and $\mathbbm{1}$ is an indicator function. Finally, $\tau$ is a temperature parameter that controls the strength of penalties on hard negative samples. Prior research demonstrated that setting a low temperature parameter helps to learn separable features, although setting it too low will result in penalizing semantically similar negative samples too heavily\cite{wang2021understanding}. We experiment with different values for $\tau$, as detailed in Section\ref{experimentalsetup}. The overall loss is bidirectional, combining the values of the two aforementioned loss functions:
\begin{equation}
    \sL = \frac{1}{\beta} \sum^{\beta}_{j=1} \Big(\ell^{(a \rightarrow v)}_{j} + \ell^{(v \rightarrow a)}_{j} \Big).
\end{equation}

\section{Experimental Setup}
\label{experimentalsetup}

\header{Datasets}
The main dataset used is a company dataset which consists of a collection of more than 90,000 music videos published on the music video platform that we collaborate with. The second dataset is the \ac{MSD}\cite{bertin2011million}. This dataset is a benchmark dataset on the task of music tagging. Note that the company dataset contains music videos, whereas the \ac{MSD} is audio only. For a detailed description and comparison of the two datasets, we refer to the GitHub repository.

\header{Evaluation method}
The two main components of our work are training the model, and evaluating the representations on downstream tasks. First, we train our model on the company dataset and experiment with several different configurations of the hyperparameters and architecture in this phase. Then, we use the representations that we learned in the first step and use them on the downstream tasks of genre classification and music tagging. This allows us to compare performance using the \ac{CL} representations to baselines from the literature.

\header{Metrics}
Following\citet{zolfaghari2021crossclr}, we use the median rank of cross-modal retrieval to evaluate our embeddings. We report both the audio-to-video median rank and the video-to-audio median rank. To evaluate performance on the downstream tasks of genre classification and music tagging, we use \ac{AUC}, in line with prior work\citet{won2020evaluation}. Additionally, we report $F_1$ score for downstream task performance.

\header{Baselines}
When evaluating the \ac{CL} representations we use the pre-trained backbone models as baselines. For the audio modality, we use \emph{musicnn}\cite{pons2019musicnn} as a baseline, and for the video modality, we use \emph{R(2+1)D}\cite{tran2018closer}. This allows us to compare our representations to representations that were not fine-tuned using \ac{CL}.

\header{Experiments} We run 3 experiments. In \textit{Experiment 1} we use \ac{CL} to learn representations for the dataset of music video segments. We start with a baseline configuration of our model, and experiment with several variations to this configuration:
\begin{enumerate*}[label=(\roman*)]
    \item \ourmodel{}: A 2-layer projection head with hidden layer size 512 and embedding size 256. We set temperature parameter $\tau$ to one.
    \item We increase the embedding size to 512.
    \item We increase the depth of the projection head to four layers. 
    \item We use a single projection head for the video modality, which projects the video embeddings to the dimensions of the \emph{musicnn} embeddings. This moves video representations closer to the audio, rather than moving both closer to each other.
    \item We lower the temperature parameter to .3, as suggested by\citet{wang2021understanding}. This penalizes hard negatives more heavily, which helps learn separable embeddings. 
    \item We train autoencoders to initialize the weights of the projection heads for the two modalities. Previous research has shown that this can improve performance on several deep learning tasks\cite{ferreira2020autoencoders, paine2014analysis}.
\end{enumerate*}

In \textit{Experiment 2}, we evaluate the representations on the downstream task of genre classification. The most used dataset in this field is the GTZAN dataset \cite{tzanetakis2002musical}, but research has pointed out that the quality of the labels in this dataset is poor \cite{palmason2017music}. This lack of a good common dataset makes it hard to compare different genre detection models and to determine what methods are the \ac{SOTA}\cite{ramirez2020machine}. In our study, we perform genre classification on the company dataset. This allows us to test our model in a real practical context, and additionally, it allows us to study the effect of the video modality on classification. We fit three different models to predict genre based on our learned representations:
\begin{enumerate*}[label=(\roman*)]
    \item $CL_{a}$: Using the contrastive audio embeddings.
    \item $CL_{v}$: Using the contrastive video embeddings.
    \item $CL_{agg}$: Using the aggregated audio and video embeddings. 
\end{enumerate*}
To evaluate the effectiveness of our learned representations, we compare these models to three baseline models, which do classification based on the pre-trained embeddings:
\begin{enumerate*}[label=(\roman*)]
    \item \emph{Musicnn}: Using the \emph{musicnn} embeddings \cite{pons2019musicnn}.
    \item \emph{R(2+1)D}: Using the \emph{R(2+1)D CNN} embeddings \cite{tran2018closer}.
    \item \emph{Musicnn+R(2+1)D}: Using the concatenated \emph{musicnn} and \emph{R(2+1)D CNN} embeddings.
\end{enumerate*}
All models are three-layer perceptrons, with hidden layers sizes of 512 and 256 respectively. 

In \textit{Experiment 3}, we evaluate the embeddings on the downstream task of music tagging. We use the same baselines and models as for genre classification. However, to facilitate reproducibility, we use a subset of the public \ac{MSD}\cite{bertin2011million} that overlaps with our company dataset. The comprehensive list of the 9416 \ac{MSD} IDs of the overlapping tracks is available on the GitHub repository referenced above. The tagging task consists of predicting several tags that were manually annotated based on the audio. These tags contain genres (e.g. ``blues''), eras (e.g. ``80s''), and moods (e.g.``happy'')\cite{pons2019musicnn, choi2016automatic}. Since our training set consists of just 9416 videos, we only use the top ten most common tags. 

\begin{figure}
 \centerline{\framebox{
 \includegraphics[width=0.9\columnwidth]{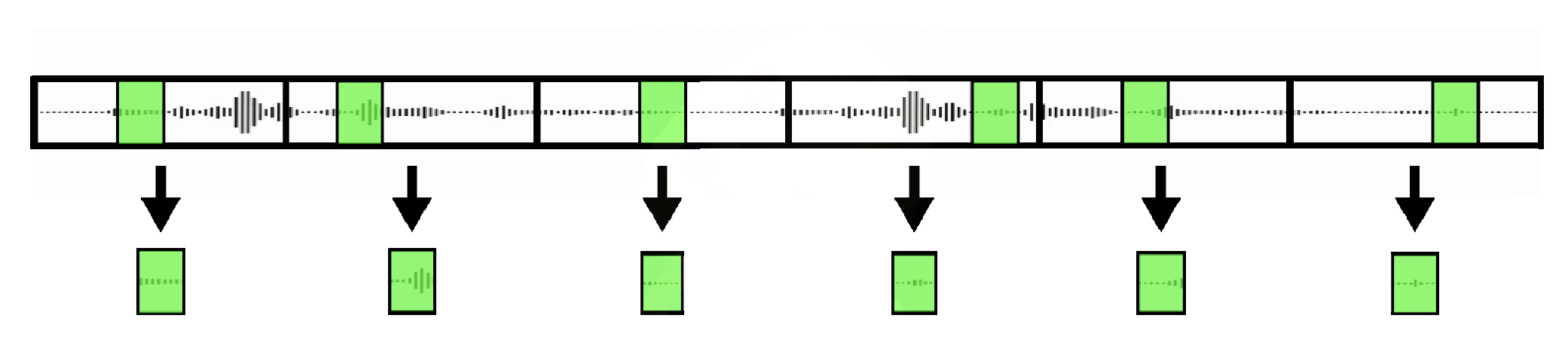}}}
 \caption{Schematic representation of the sampling procedure.}
 \label{fig:sampling}
\end{figure}

\header{Implementation Details}
In order to run our experiments in relatively quick succession, we stored just six five-second non-overlap-\\ping segments per music video, rather than the entire music video. We sampled the segments by dividing the track up into six equally sized sections and sampling a random five-second segment from each section, preventing overlap between the segments, this procedure is represented schematically in Figure\ref{fig:sampling}.

Consecutive video frames have high redundancy \cite{wang2016temporal}. So in line with previous work \cite{simonyan2014two, carreira2017quo} we choose to sample a random set of frames with equal temporal spacing between them. We save 50 frames at a resolution of 768x432. The corresponding audio was saved at a sampling rate of 16,000. Data was split up into a training set of 80 percent, and a validation and testing set of 10 percent each. 

Separate segments of the same music video were constrained not to appear in the same batch. In We applied random cropping to 112x112 in order to get the correct input shape for the network, 
 using resizing to prevent the cropped videos to contain an unreasonably small fraction of the pixels in the original video, which is a common data augmentation in visual deep learning \cite{takahashi2018ricap}.

We use a batch size of 1000, with an initial learning rate of .01, and exponential learning decay with a gamma parameter of .95, following \citet{li2019exponential}. We implement an early stopping rule, stopping when the validation loss does not decrease over a period of three epochs. 

%% !TEX root = ../main.tex

\section{Results}
\label{sec:results}

% \begin{figure}[b]
%   \centering
%   \subfloat[Loss function]{\includegraphics[width=0.4\textwidth]{img/train_val_loss.png}\label{metrics:f1}}
%   \hfill
%   \subfloat[Median rank]{\includegraphics[width=0.4\textwidth]{img/medianranks.png}\label{metrics:f2}}
% \caption{Training and validation metrics during contrastive training of the baseline model. The loss function shows a slight decrease whereas the median rank does not drop below the chance level consistently.}
%  \label{train_metrics}
% \end{figure}

\header{Experiment 1}
In the first experiment, we explore different configurations of our approach. Table\ref{experiments} presents the results of the experiment. The median rank does not change across different selected configurations.

\begin{table}[]
\centering
\caption{Results of Experiment 1. The best performance is highlighted in
bold.}
\label{experiments}

\begin{tabular}{lcc}
\toprule
\textbf{Configuration} & \textbf{$MR_{a\rightarrow v}$} & \textbf{$MR_{v\rightarrow a}$}  \\            
\midrule
\ourmodel{}                   & 507 & 502 \\
Embedding size 512          & \textbf{494} & 498\\
Four projection layers      & 503 & 501\\
Single projection head      & 499 & \textbf{497}\\
$\tau = 0.3$             & 497 & 503\\
Autoencoder initiated       & 501 & 500\\
\bottomrule
\end{tabular}

\end{table}

\header{Experiment 2 and 3}
Table\ref{tag_results} shows the average area under the curve and F1 score for each baseline and model on the downstream task of music tagging. These results present two main findings. First, the networks that are trained on embeddings that include information about audio clearly outperform models that are only based on a video. Second, the performance of the baseline models is better than the performance of the models that were fine-tuned using \ac{CL}. Given the failure to unite embeddings from the different modalities, these results are less surprising. 

The results for the downstream task of genre classification are similar to the results for music tagging. Table\ref{tag_results} presents the same two statistics for the task of genre classification. Similar to the results above, we see thfat models including audio data outperform models that are solely based on a video, and more importantly, that all three baselines outperform their \ac{CL} counterparts. 

%\input{tables/tagging_results}
%\input{tables/genre_results}
%\input{ecir_23_music_video_reco/tables/test}

% \usepackage{booktabs}
% \usepackage{multirow}
\begin{table}[]
\centering
\caption{Results of Experiments 2 and 3. The best performance is highlighted in
bold.}
\label{tag_results}
\begin{tabular}{lcccc}
\toprule
\multirow{2}{0cm}{\textbf{Model}} & \multicolumn{2}{c}{\textbf{Music tagging}}                     & \multicolumn{2}{c}{\textbf{Genre classification}} \\
\cmidrule(l){2-3} \cmidrule(l){4-5}
                                         & \multicolumn{1}{c}{\textbf{AUC}} & \multicolumn{1}{c}{\textbf{$F_1$}} & \multicolumn{1}{c}{\textbf{AUC}} & \multicolumn{1}{c}{\textbf{$F_1$}} \\
\midrule
Musicnn                & 0.78 & 0.22 & 0.79 & 0.18 \\
R(2+1)D            & 0.66 & 0.09 & 0.68 & 0.05 \\
Musicnn+R(2+1)D          & \textbf{0.78} & \textbf{0.24} & \textbf{0.82} & \textbf{0.21} \\
$CL_{a}$      & 0.70 & 0.10 & 0.69 & 0.08 \\
$CL_{v}$      & 0.61 & 0.09 & 0.63 & 0.04 \\
$CL_{agg}$ & 0.71 & 0.11 & 0.68 & 0.07 \\
\bottomrule
\end{tabular}

\end{table}

%% !TEX root = ../main.tex

\section{Qualitative analysis of representations}
\label{sec:qualtatie-analysis}

To better understand our results, we analyzed our representations qualitatively on the level of entire music videos as well as individual samples.
    
\header{Music videos} 
We selected 25 random seed videos and retrieved the three most similar music videos to each seed, based on the embeddings aggregated over the audio and video modalities. In general, the most similar music videos did not appear to be similar to the seed video at all in terms of both audio and video. One exception to this rule was live performances, which are often most similar to other live performances. When taking into account only the video modality, rather than the aggregate over modalities, some other patterns were also present: Some seed videos that were all black and white, as well as some videos of people dancing or playing instruments, would result in similar videos being retrieved. One thing that these exceptions have in common, is that they are very consistent in terms of video across different parts of the music video. This could point to the fact that aggregating representations of different segments of the same music video is not a valid way to get a representation of the music video, especially when different parts of the music video are very inconsistent. This is particularly often the case for the video modality, as music videos often consist of a wide variety of shots and scenes. 

\header{Segments} We also examined similarity on the level of individual segments. We took the same approach as before, this time selecting 20 random seed segments, and retrieving the three most similar segments to this seed segment. As expected, the results were better for single segments. There were more cases in which the retrieved videos were similar to the seed, although in general the retrieved videos were still not similar. For example, some seeds for which the retrieved segments were similar were close-ups of women singing, shots of people playing guitar, segments with a dark background with bright neon accents, and shots of men talking. Overall similar segment retrieval was poor, but there were more exceptions in which there was some specific similarity between the videos. 

\section{Discussion and limitations}

Contrastive learning of audiovisual representations was not able to generalize to the field of music videos. The contrastive audiovisual representations did not improve performance on either of the downstream tasks, answering RQ1. Additionally, with respect to RQ2, results indicated that our representations can not be used to infer music video similarity. In this section, we discuss some of the potential reasons for these issues, as well as some limitations to our approach. 

\header{Discussion} The main challenge for multimodal learning consists of bridging the heterogeneity gap\cite{guo2019deep}. \ac{CL} is one approach to this problem, and some notable applications involve \ac{CL} for images and their captions\cite{radford2021learning}, videos and their descriptions\cite{zolfaghari2021crossclr} and videos and their audio streams \cite{wang2021multimodal}. Compared to our use case, the heterogeneity gap in these studies is small. An image and its caption are closely related, as the caption often describes the image. The two modalities both refer to the same concept. This is also true for videos and their descriptions, or videos of an event and the sound it produces. However, the relationship between music and its video is less direct. As opposed to the other examples, the two modalities are not both directly referring to a single concept. Rather than describing the music, the video modality serves as an extra means to entertain the user and convey emotion. This large heterogeneity gap is reflected in our qualitative analysis of similarity: The music network encodes for features relating to the types of instruments used, and the video network encodes for features like the color of the video as well as specific actions like singing or dancing. The relationship between these features is not always immediately apparent. Black and white videos vary greatly in the type of instruments used, and the fact that a music video contains people dancing, does not tell much about the instruments played. This is different from the video-audio use case from previous research, where different videos of similar events are likely to have similar audio streams since the event in the video is often the source of the audio \cite{wang2021multimodal}. We believe that the fact that the two modalities in music videos are less tightly connected is an important cause for the failure to generalize contrastive learning to the music-video domain. 

\header{Limitations} 
% The main limitation of our research is that the training process fails to pull embeddings for corresponding audio and video segments closer to each other. This is a serious problem for our project, as uniting embeddings from different modalities is the fundamental idea of \ac{CL}. A possible explanation for this result is that the heterogeneity gap is too big in our use case, making it hard to represent audio and video in the same space. 
One limitation of our approach concerns the input size of the encoder networks. Due to the high dimensionality and complexity of the input data, we were limited to training on segments of only three seconds. This makes the task of retrieving corresponding segments from the other modality significantly harder. Even for a human, retrieving the corresponding piece of music based on a three-second video clip is difficult. Additionally, this short input length makes it harder to compute the similarity between complete music videos, which is necessary for \ac{CBR}. We tried aggregating representations of segments to come to a single representation per music video, but our qualitative analysis of similarity revealed that this aggregation degraded the performance, especially so when the input data was inconsistent across segments, which is usually the case for video data from music videos. It might be more fruitful to use a larger input size or to extract features that are more consistent across the duration of the music video. 

% !TEX root = ../main.tex

\section{Conclusions}
\label{sec:conclusions}
In this work, we examined several \ac{CL} bi-encoder architectures and explored their generalizability towards the music video domain. We discovered that the models struggled with 
A contrastive deep learning model was not able to unite embeddings from the audio and video modalities in music videos. These results demonstrate that aligning video and audio data is an open and challenging problem.

Given the large heterogeneity gap, future research should look into multimodal alternatives to \ac{CL} that rely less heavily on the association between features in different modalities. Multimodal fusion might for example be more suitable, as these models combine features from different modalities based on a supervised task, rather than solely based on the relationship between features in the different modalities\cite{nojavanasghari2016deep}. 

Additionally, future research could focus on using encoder networks that learn features that are more consistent over time.  Most current research on video processing is based on short video clips of single events, faces, or poses \cite{ren2019survey, santos2019deep}. However, ideally, the video encoder would learn features relating to the style of the video. Potentially, models concerned with style transfer for video data \cite{chen2017coherent, li2019learning} could learn features that are more closely related to the audio of the song.

\bibliographystyle{ACM-Reference-Format}
\bibliography{bibliography}
\end{document}